# Plattformen und neue Technologien im Journalismus

Ergebnisse einer Online-Befragung von Journalistinnen und Journalisten in Deutschland

Benjamin Rech[a] und Matthias Meyer[b]

[a] Ostfalia Hochschule für angewandte Wissenschaften, Institut für Medienmanagement, Salzgitter, Deutschland

[b] Technische Universität Chemnitz, Informatik für Geistes- und Sozialwissenschaftler, Chemnitz, Deutschland

**Abstract**

In an online survey in December 2020, 385 journalists in Germany were surveyed about platforms in journalism and about their frequency of use and willingness to adopt emerging technologies. Journalists have a commitment to publish on a journalism platform on a full-time basis. Freelancers have a higher commitment than employed journalists. A platform subscription model is rated more attractive than advertising for a platform. Employed journalists on the other hand consider advertising more attractive than freelance journalists. For German journalists it is important that the platform is developed in Europe or Germany and that it sets high standards on data protection. Multimedia forms and interactive elements are used occasionally, often or always. Stories or Reels are predominantly not used. AI software as well as editorial analytics are rarely or never used. Apart from stories or reels, journalists intend to use multimedia forms and interactive elements more often in the future. They are receptive to software for research process documentation as well as to the analysis of indicators of their own publications. Software as a support for text production, image selection or headline suggestions is mostly rejected.

In einer Online-Befragung wurden im Dezember 2020 385 Journalistinnen und Journalisten in Deutschland zu Plattformen im Journalismus und zu ihrer Nutzungshäufigkeit sowie Nutzungsbereitschaft neuer Technologien befragt. Journalistinnen und Journalisten haben eine Bereitschaft auf einer Journalismusplattform hauptberuflich zu publizieren. Freie haben eine höhere Bereitschaft als Angestellte. Ein Plattform-Abonnement wird attraktiver bewertet als die Werbefinanzierung einer Plattform. Es ist Journalistinnen und Journalisten wichtig, dass die Plattform aus Europa oder Deutschland stammt und besonderen Wert auf Datenschutz legt. Multimediale Darstellungsformen und interaktive Elemente werden gelegentlich, oft oder immer genutzt. Aneinandergereihte Bilder und Videos (Storys oder Reels) werden überwiegend nicht genutzt. „KI"-Software als Produktionshilfe sowie Analyse-Tools zur Auswertung eigener Veröffentlichungen werden selten oder nie genutzt. Mit Ausnahme von Storys oder Reels intendieren Journalistinnen und Journalisten multimediale Darstellungsformen und interaktive Elemente in Zukunft häufiger zu nutzen. Gegenüber Software zur Dokumentation und Darstellung des Rechercheprozesses sowie zur Analyse von Kennzahlen eigener Veröffentlichungen sind sie aufgeschlossen. Software als Unterstützung zur Textproduktion, Bildauswahl oder für Überschriftenvorschläge wird im Vergleich kritisch gesehen.

**Schlagworte**: Journalismusplattform, Journalistische Plattformen, Digitaler Journalismus, Medienmanagement, Quantitative Befragung

# Inhaltsverzeichnis

# 1 Einleitung in die Turbulenzen des (digitalen) Journalismus in Deutschland

**Das Geschäftsmodell des analogen in Redaktionen organsierten Journalismus, mit seinem zweiseitigen – über Kunden und Werbung – finanzierten Erlösmodell, scheitert im Internet:**

Die Einnahmen der Redaktionen durch Werbung und Vertrieb sinken. 2019 konnte ein Verlust der jährlichen Werbeerlöse von 21,81 Prozent (612 Mio. Euro) bei allen Tageszeitungen sowie Wochen- und Sonntagszeitungen im Vergleich zu 2015 festgestellt werden. Die Erlöse aus dem Vertrieb (Abo und Einzelverkauf) bleiben stabil, aufgrund erhöhter Verkaufspreise. So wurde ein Abonnement einer täglich erscheinenden Lokal- und Regionalzeitung 2019 im Vergleich zu 2015 monatlich durchschnittlich 6,89 Euro (inflationsbereinigt: 5,30 €) teurer. (Keller und Stavenhagen 2020) Die verkaufte Gesamtauflage der Tageszeitungen in Deutschland sinkt konstant. Im Jahr 2020 wurden täglich 12,5 Millionen Exemplare verkauft, drei Millionen weniger als noch 2015. Seit 1995 hat sich die verkaufte Auflage damit halbiert. (BDZV 2020)

Gleichzeitig steigt die Nutzung digitaler Medien zur Information. 34 Prozent aller Menschen in Deutschland lesen im Jahr 2020 aktuelle Nachrichten im Internet (2019: 7 Prozentpunkte weniger). 94 Prozent der deutschsprachigen Bevölkerung ab 14 Jahren nutzen inzwischen zumindest gelegentlich das Internet. (Beisch und Schäfer 2020)

Trotz steigender Nutzung des Internets und steigender Zahlungsbereitschaft für digitale Premium-Inhalte (Deloitte 2020), besteht nur eine geringe Zahlungsbereitschaft für die digitalen Produkte der Redaktionen: Im Digital News Report des Reuters Institute geben zehn Prozent der Befragten in Deutschland an, im Jahr 2020 für Online-Nachrichten Geld bezahlt zu haben (Hölig und Hasebrink 2020). Laut einer Befragung von PWC (2019) haben 39 Prozent der 18- bis 29-Jährigen bereits für Online-Journalismus Geld ausgegeben, wohingegen 59 Prozent der Deutschen noch nie für die Berichterstattung online gezahlt haben. 54 Prozent nennen eine fehlende inhaltliche Relevanz als Grund, nicht für Online-Journalismus bezahlen zu wollen. Das Konzept der Gratis-Mentalität (bei Rezipienten wie auch in Reaktion bei Redaktionen) kann 40 Prozent der Varianz mangelnder Zahlungsbereitschaft für digitalen Journalismus in Deutschland erklären (O'Brien et al. 2020).

Lobigs (2018, S. 301–305) betitelt den digitalen Werbemarkt für Journalismus im Internet als „kaputt". Der Tausend-Kontakte-Preis ist online „bereits im Jahr 2012 gerade einmal bei zehn Prozent der Print-Pendants" (Kansky, S. 151–152), und unterliegt seit dem einem ungebremsten Preisverfall". Neben der Verdrängung durch Technologiekonzerne, die aufgrund ihrer Datenhoheit im Bereich Search- und Programmatic Advertising eine Konkurrenz zu klassischen Werbemodellen darstellen, nennt Lobigs die Verbreitung von Ad-Blockern, das Überangebot beim Display-Ad-Inventar, die zunehmende Smartphone-Nutzung sowie die Angebotsausweitung bei alternativen Werbeformen wie dem Content Marketing als Gründe für das Scheitern des Werbeerlösmodells für den Journalismus im Internet. Alternative und lukrativere Werbeformate wie Native Advertising – insofern sie denn als solche wahrgenommen werden – unterminieren das Vertrauen Rezipierender in den Verlag (Amazeen und Wojdynski 2020) .

Spätestens 2025 sollen die digitalen Vertriebsumsätze die Rückgänge des klassischen Anzeigengeschäfts kompensieren (Schickler und BDZV 2020). Geringe Einnahmen durch digitale Abonnements und der „kaputte" digitale Werbemarkt (Lobigs 2018) können die sinkenden Vertriebserlöse der gedruckten Zeitung bisher allerdings nicht ausgleichen, weshalb eine Querfinanzierung der Digital-Angebote in Zukunft als unwahrscheinlich anzusehen ist.

Der Journalismus sieht sich als gesellschaftliche Institution einer Verdrängung ausgesetzt (Lobigs 2018). Substituierende Produkte müssen – im Gegensatz zu redaktionellem Journalismus – keine Mehrwertkommunikation für ihre Legitimation betreiben – bildhaft gesprochen: Mit Google findet man schnell fundierte Informationen, bei Twitter ist man sofort informiert, bei Instagram näher an Personen des öffentlichen Lebens und Facebook zeigt einem genau die Inhalte an, die die eigenen Freunde lesen. Auf sozialen Netzwerkseiten stehen Journalistinnen und Journalisten in direkter Konkurrenz um Reichweite mit allen aktiven Nutzerinnen und Nutzern der Plattformen. Angesichts dieser aufmerksamkeitsökonomisch induzierten Journalismus-Krise (Rau 2014) braucht es ein meritorisches Gesamtkonzept für eine zukunftsfähige Ökonomie des Journalismus (Rau 2020) und damit auch zur Legitimation von Journalismus als „vierte Gewalt".

Ist also möglicherweise eine eigene Plattform für Journalismus die Lösung? Braucht es ein digitales Presse-Grosso oder einen öffentlich-rechtlichen Journalismus? Im Sammelwerk "Money for Nothing and Content for Free?" von Wellbrock und Buschow (2020) analysieren die Autorinnen und Autoren umfangreich verschiedene Perspektiven von Paid Content, Plattformen und der Zahlungsbereitschaft von Nutzerinnen und Nutzern für digitalen Journalismus.

Unsere Studie ergänzt die aktuelle Forschung zu journalistischen Plattformen um die Kommunikatorperspektive und damit um den zentralen Akteur im Journalismus: Die Journalistin. Den Journalisten. Die zugrundeliegenden Forschungsfragen lauten:

*FF1: Haben hauptberufliche Journalistinnen und Journalisten in Deutschland eine Bereitschaft ihre Beiträge auf einer Plattform zu veröffentlichen, insofern diese den eigenen Lebensunterhalt sichert?*

*FF3: Welche Finanzierungsmodelle für eine solche Plattform empfinden Journalistinnen und Journalisten als attraktiv und welche Anforderungen stellen sie?*

*FF2: Wie hoch sind Nutzungshäufigkeit und Nutzungsintention digitaler Technologien durch Journalistinnen und Journalisten?*

# 2 Theorie

Der folgende Forschungsstand beleuchtet Plattformen im Journalismus sowie technologische Innovationen in der Branche und für die journalistische Arbeit. Die theoretische Vorarbeit mündet in eine Operationalisierung der Konstrukte und damit zu Hypothesen.

## 2.1 Plattformen im Journalismus und eine „Journalismusplattform"

Eine Plattform ist aus wirtschaftswissenschaftlicher Sicht einer von mehreren möglichen Intermediären in einem mindestens zweiseitigen Markt (engl. Two-Sided-Market). "Two-sided (or more generally multi-sided) markets are roughly defined as markets in which one or several platforms enable interactions between end-users, and try to get the two (or multiple) sides 'on board' by appropriately charging each side." (Rochet und Tirole 2006, S. 645)

Rysman (2009, S. 125) spezifiziert, dass in einem zweiseitigen Markt "1) two sets of agents interact through an intermediary or platform, and 2) the decisions of each set of agents affects the outcomes of the other set of agents, typically through an externality".

Eine sektorale Plattform ist im Gegensatz zu generalistischen Plattformen branchenspezifisch (van Dijck et al. 2018). „Journalismusplattform" ist dahingehend ein verkürzter Begriff für eine sektorale Plattform für Journalismus[1]. Für diese Studie wird als Journalismusplattform einer von mehreren möglichen digitalen Intermediären verstanden, der zwischen Anbietermarkt (Journalisten und Redaktion) und Konsumentenmarkt (Rezipienten) monetäre Transaktionen ermöglicht und bei der das Plattformnutzungsverhalten (Transaktionen und Interaktionen) der einen Gruppe reziprok die andere Gruppe beeinflusst.

Nach der zuvor getroffenen Definition des Begriffs Journalismusplattform gelten aktuell Amazon Prime Reading, Apple News +, Cafeyn, inkl sowie Readly als Journalismusplattformen. Zudem ist davon auszugehen, dass sich die beiden Startups Publikum und NewsAdoo vom Geschäftsmodell Nachrichten-Aggregator hin zur Plattform entwickeln werden. Als Nachrichten-Aggregator, der auch als Feed-Aggregator, Feed-Reader, News-Reader, RSS-Reader oder einfach als Aggregator bezeichnet wird, wird eine Anwendungssoftware oder eine Web-Anwendung verstanden, die syndizierte Web-Inhalte wie Online-Zeitungen, Blogs, Podcasts und Video-Blogs an einem Ort zusammenfasst. Neben Publikum und NewsAdoo können Bundle, Feedly, Flipboard, Google News, Nuzzera, Rivva und Squid als Nachrichten-Aggregatoren identifiziert werden. Flattr, Patreon und Steady dagegen sind Social Payment Services. Substack und Revue gelten als Newsletter-Services. Die sozialen Netzwerkseiten Facebook,

---

[1] Journalismus wird in dieser Studie nach dem normativ-pragmatischen Ansatz als kommunikatives Handeln (Bucher 2004) definiert, dessen Ziel die "gelingende gesellschaftliche Kommunikation" ist. Gesellschaftliche Kommunikation "gelingt dann (dort), wenn (wo) der Journalismus eine mediale Wirklichkeit erzeugt, die von den Kommunikationspartnern (Akteuren und Rezipienten) als Orientierung über aktuelle Ereigniszusammenhänge genutzt, zumindest so verstanden wird." (Haller 2003, S. 181)

Twitter und YouTube sind als sektorübergreifende, generalistische Plattformen einzuordnen und ermöglichen bisher nur Aufmerksamkeitstransaktionen.

In der Medienwirtschaft wird zudem von einer Plattformisierung gesprochen (Eisenegger et al. 2021; Jarren 2019; van Dijck et al. 2018; Nieborg et al. 2019). Plattformisierung beschreibe den Prozess der fortschreitenden Durchdringung medialer Kommunikation durch Plattform-Geschäftsmodelle und damit einhergehender Veränderungen der Geschäftsmodelle von Medienorganisationen, Veränderungen bei der Relevanzbestimmung, Erstellung und Distribution von Medienprodukten sowie bei der Hoheit über ihren Darstellungskontext.

Die Frage nach einem Spotify oder Netflix für Journalismus wird aufgrund ihrer Hypothetik überwiegend durch und in journalistischen – oder meta-journalistischen – Medien diskutiert (Baekdal 2021; Buhre 2021; Schade 2021; Graf 2020). Eine erste wissenschaftliche, multiperspektivische Annäherung an die „Plattform-Frage" veröffentlichen Wellbrock und Buschow (2020) mit ihrem Sammelband „Money for nothing, Content for Free?". Auf Basis von Fokusgruppen mit 55 deutschen Nachrichtennutzerinnen und -nutzern analysieren Buschow und Wellbrock (2020) erwartete Merkmale abonnementbasierter, anbieterübergreifender Plattformen im Journalismus. Die Befragten äußern den Wunsch nach „hochwertigen, exklusiven und vielfältigen journalistischen Inhalten", gleichzeitig sehen sie die Probleme bestehender Nachrichten-Aggregatoren oder sozialer Netzwerkseiten: Eine inhaltliche Redundanz und damit einhergehende Informationsüberflutung, der hohe eigene Selektionsaufwand sowie eine durch die Plattform vorselektierte Nachrichtenauswahl und / oder eine zeitlich verzögerte Bereitstellung. Hinsichtlich Benutzerfreundlichkeit, Design und Komplexität wünschten sich Nachrichtennutzerinnen und -nutzer eine einzige Plattform, auf der sie Inhalte mehrerer Redaktionen und freier Journalistinnen und Journalisten lesen können. Um den Konsum journalistischer Inhalte angenehm zu gestalten, nennen die Befragten Funktionen wie das Abonnieren einzelner Journalistinnen und Journalisten, eine Merkliste, ein Archiv sowie personalisierte Empfehlungen. Personalisierte Empfehlungen werden von den Befragten kritisch diskutiert. Einerseits werden Empfehlungs- und Sortieralgorithmen hinsichtlich ihrer Komplexitätsreduktion positiv gesehen, andererseits werden negative Auswirkungen auf die Meinungsbildung durch Überselektion befürchtet. Um Vertrauen zu einer Plattform für Journalismus aufbauen zu können, wünschten sich Nachrichtennutzerinnen und -nutzer Transparenz und Unvoreingenommenheit bei der Auswahl teilnehmender Redaktionen und Journalisten. Ob man auf einer solchen Plattform einzelne Redaktionen auswählen und abonnieren kann oder eine themenbasierte (entbündelte) Zusammenstellung von Inhalten angeboten bekommt, sehen die Teilnehmerinnen und Teilnehmer der Diskussion unterschiedlich. Bei der Diskussion hinsichtlich der preislichen Gestaltung einer Journalismusplattform orientieren sich die Befragten an den Modellen etablierter Plattformen wie Netflix und Spotify: Ein günstiges monatliches Abonnement, keine Werbung, kurze Kündigungsfristen und die Möglichkeit den Account mit Freunden oder der Familie zu nutzen. (vgl. für eine Übersicht: Buschow und Wellbrock 2020, S. 138)

Welche Präferenzen Konsumentinnen und Konsumenten in Deutschland für verschiedene Finanzierungsmodelle haben, erheben die Autoren in einer quantitativen Panel-Befragung (n = 1000) für die Modelle Einzelartikel, Einzel-Abonnement, Plattform und Spende (Wellbrock 2020a, S. 171). Die Bezahlabsicht sei für alle Modelle im Mittel zwischen 2,08 und 2,23 Punkten, was knapp über der Skalenausprägung „unwahrscheinlich" bei Punktwert 2 liegt. Den höchsten Wert erreiche in dieser Frage nach der zukünftigen Bezahlwahrscheinlichkeit das Plattform-Modell, den niedrigsten das Modell „Einzelartikel". Auch die Kaufabsicht liege im Mittel bei allen Modellen zwischen 2,30 und 2,57, was zwischen den Skalenausprägungen „trifft nicht zu" (2) und „trifft eher nicht zu" (3) liegt. Das Spendenmodell erreiche in dieser Frage nach der zukünftigen Kaufwahrscheinlichkeit den höchsten Wert, das Abonnement eines digitaljournalistischen Produktes den niedrigsten (Plattform MW = 2,48).

## 2.2 Technologische Innovationen im Journalismus

Die journalistischen Verlage in Deutschland sind mit fehlenden Strategien zur Personalisierung ihrer digitalen Angebote "Zaungäste der Plattform-Tektonik" (Lobigs und Mündges 2020). Es fehlen technologische Wege, Journalismus als gesellschaftliches Konzept in Empfehlungs- und Sortieralgorithmen für Nachrichten - respektive auf einer Plattform - abzubilden. Die internationale Forschung zu Empfehlungs- und Sortieralgorithmen in der Nachrichtendomäne (engl. News Recommender Systems) ist sich der gesellschaftlichen Implikationen zwar gewahr, konkret:

- der Fragmentierung von Öffentlichkeit durch und die demokratische Rolle von Empfehlungs- und Sortieralgorithmen für Nachrichten (Helberger 2019)
- dem Konzept von Filterblasen nach Pariser (2011)
- der metaphorischen Echokammer, genauer den Phänomenen „Selective Exposure" (Garrett 2009) und der kognitiven Dissonanz (Bright et al. 2020)
- der Verweigerung kognitiver Dissonanz bei gegensätzlichen Meinungen (Beam 2014)
- der politischen Meinungspolarisierung und der Assimilation von Algorithmen-Bias (Dandekar et al. 2013)

Allerdings ist die Frage nach einem verantwortungsvollen Empfehlungs- und Sortieralgorithmus ein komplexes informatisches Problem und die betroffenen Wissenschaftsdisziplinen arbeiten nur in Ausnahmen (Ocaña und Opdahl 2020) interdisziplinär zusammen. In der Journalismus-Branche lässt sich ein sogenannter „Quantitative Turn" beobachten, welcher unter den Begriffen der computergestützten Berichterstattung, dem Datenjournalismus und dem sog. Roboterjournalismus (auch automatisierter Journalismus, algorithmisierter Journalismus) wissenschaftlich untersucht werden (Coddington 2015). Maschinelles Lernen kann in Anwendungen für Journalisten konkret Sätze vervollständigen oder ganze Texte schreiben (GPT-3: Brown et al. 2020), Überschriftenvorschläge machen (Gu et al. 2020), eine automatische Bildauswahl treffen (Liu et al. 2020) oder geschriebenen Text stilistisch optimieren (). In einer internationalen Befragung von Nachrichtenorganisationen zum Thema maschinellem Lernen nannten Verantwortliche drei Hauptmotive für den Einsatz von KI: Die Arbeit der Journalisten effizienter zu gestalten (68 Prozent der Antworten), den Nutzern relevantere Inhalte zu liefern (45 Prozent), die Wirtschaftlichkeit zu steigern (18 Prozent). Als größte Herausforderungen bei der Einführung von maschinellem Lernen nannten die Befragten finanzielle Ressourcen (27 Prozent) und Wissen oder Fähigkeiten (24 Prozent). Neben diesen „einfachen Herausforderungen" wurde eine „cultural resistance" (24 Prozent) angeführt, einschließlich der Angst vor dem Verlust des Arbeitsplatzes, vor einer Änderung der Arbeitsgewohnheiten und einer allgemeinen Ablehnungshaltung gegenüber neuen Technologien. Mangelndes Wissen über maschinelles Lernen in der gesamten Redaktion und ein Mangel an strategischem Managementwissen machen 36 Prozent der Antworten aus. (Beckett 2019)

Neben technologischen Fortschritten im Bereich des maschinellen Lernens haben sich – getrieben durch die veränderte insbesondere mobile Medienrezeption, durch Trends in sozialen Netzwerken und durch Lösungsansätze strategischer Kommunikation (Hooffacker und Wolf 2017) – neue Darstellungsformen für digitalen Journalismus entwickelt. Insbesondere ist hier das multimediale Storytelling als Konzept zu erwähnen, welches Medienformen wie Fotos, Videos, Animationen, Audio-Elemente, interaktive Grafiken, Datenvisualisierungen, 360°- Fotos, 360°-Videos und VR-Elemente nutzt um Geschichten ansprechender zu erzählen (Radü 2019b). Daneben ist die Kontextualisierung von Diskursen im Internet durch die Einbettung von Social-Media-Beiträgen ein zentrales Element digitaler Nachrichtenformate (Thorsen und Jackson 2018).

Da auch digitaler Journalismus den aufmerksamkeitsökonomischen Grundprinzipien des Internets folgen muss, werden Nutzerinteraktionen mit journalistischen Inhalten gemessen und auf diverse Kennzahlen (Klicks, Verweildauer, Anzahl gesehener Werbebanner, etc.) optimiert (Sehl und Eder 2020). Das Konzept dieser Nutzerzentrierung wird in der Wissenschaft überwiegend kritisch gesehen (Ellers 2020).

## 2.3 Operationalisierung theoretischer Vorarbeit

Der in den vorangegangenen Kapiteln dargestellte Forschungsstand mündet in eine Operationalisierung (siehe Tab. 1). Soziodemografika sowie Aspekte der journalistischen Tätigkeit bilden den Rahmen um die zentralen Konstrukte der Plattformbereitschaft, der Plattform-Finanzierung, normativer als auch inhaltlicher Anforderungen, der Nutzungshäufigkeit und Nutzungsintention multimedialer und interaktiver Darstellungsformen sowie der Nutzungshäufigkeit und Nutzungsintention von Software im journalistischen Arbeitsprozess.

Tab. 1: Operationalisierung der Konstrukte

| Konstrukte | Variablen | Quellen |
|---|---|---|
| 1 Journalistische Tätigkeit | Redaktionsgröße, Medienformen, Arbeitsverhältnis, Berufstätigkeit in Jahren, Wöchentliche Arbeitszeit, Vergleich der wöchentlichen Arbeitszeit, Anzahl von Veröffentlichungen | Weischenberg et al. (2006) Gautschi (2016) Steindl et al. (2017) |
| 2A Plattformbereitschaft | Hauptberufliche Plattformnutzungsbereitschaft, Nebenberufliche Plattformnutzungsbereitschaft | - |
| 2B Plattform-Finanzierung | Attraktivität von: Micro Payments, Werbung, Klubfinanzierung, Monatliches Plattform-Abonnement | Wellbrock (2020a) |
| 3A Normative Anforderungen | Verlagsplattform, Öffentlich-rechtliche Plattform, Plattform aus Europa, Plattform mit hohem Datenschutz, Plattform ohne Tech-Giganten | Wellbrock (2020b) |
| 3B Inhaltliche Anforderungen | Algorithmische Relevanzbestimmung, Algorithmische Qualitätsbewertung, Algorithmentransparenz, Medienmarke oder persönliche Marke, One-Stop-Shop, Veröffentlichungsrestriktionen | Buschow und Wellbrock (2020, S. 138) |
| 4A Multimedia | Nutzungshäufigkeit und Nutzungsintention von Storys, interaktiven Elementen und einer multimedialen Content-Darstellung | Radü (2019a) |
| 4B Software | Nutzungshäufigkeit und Nutzungsintention von Software als Produktionshilfe, als Recherche-Tool und als Analyse-Tool | Beckett (2019) Loosen und Solbach (2020) Sehl und Eder (2020) |
| 5 Soziodemografika | Einkommen, Geschlecht, Alter, Bildung, | Beckmann et al. (2016) Steindl et al. (2017) |

# 3 Methode: Online-Befragung von Journalisten in Deutschland

Um herauszufinden ob bei Journalistinnen und Journalisten in Deutschland eine hauptberufliche Nutzungsbereitschaft für eine Journalismusplattform besteht, welche Anforderungen diese an eine solche Plattform haben und wie hoch die Nutzungshäufigkeit und -intention technologischer Innovationen ist, wurde zwischen dem 2. Dezember und dem 31. Dezember 2020 eine quantitative Studie durchgeführt.

**Grundgesamtheit hauptberuflicher deutscher Journalistinnen und Journalisten**

Die Grundgesamtheit hauptberuflicher deutscher Journalistinnen und Journalisten ist zum Zeitpunkt dieser Studie nicht hinreichend bekannt (Tab. 2) bzw. nicht korrekt ermittelbar aufgrund einer unklaren Berufsdefinition (Malik 2011). Für 2017 erfasst die Bundesagentur für Arbeit (2019) 129.800 erwerbstätige Redakteure und Journalisten im Bereich Publizismus. Die methodisch beste Schätzung auf Basis redaktioneller Einheiten stammt von Steindl et al. (2017) auf Basis der Studie von Weischenberg et al.

(2006). Die Autorinnen und Autoren gehen für das Jahr 2015 von einer Grundgesamtheit von 42.500 Personen (2005: 48000) aus.

Tab. 2: Studien zu Journalistinnen und Journalisten in Deutschland

| **Autor** | **Anteil M** | **Anteil W** | **Anteil Freie** | **Alter** | **GG** |
|---|---|---|---|---|---|
| Weischenberg et al. (2006) | 62,70% | 37,30% | 25,18% | < 25: 2,7 %; 26-35: 29,7 %; 36-45: 39,6 %; 46-55: 21 %; 56-65: 6,5 %; > 65: 0,5 % | 48.000 |
| Kaiser (2012) | - | - | 35,86% | - | 72.500 |
| Steindl et al. (2017) | 60,90% | 39,10% | 17,60% | MW = 45,58; SD = 10,5 | 41.250 |
| Gautschi (2016) | 62,04% | 37,96% | 31,60% | MW = 49,91; SD = 10,23 | 29.826 |
| Bundesagentur für Arbeit (2019) | - | - | - | - | 129.800 |
| Diese Studie (Rohdaten, 2020) | 53,00% | 46,90% | 39,1 % | MW = 40; SD = 12,91 | 5.705 |
| Diese Studie (gewichtet, 2020) | 63,8 % | 36,2 % | 68,4 % | < 29: 8,2 %; 30-39: 19,9 %; 40-49: 18 %; > 50: 54 % | 30.000 |
| Hanitzsch und Rick (2021) | 61,70% | 37,70% | 61,10% | < 29: 8,4 %; 30-39: 19,8 %; 40-49: 16,8 %; > 50: 55 % | 30.000 |

**Auswahlverfahren und Stichprobenkonstruktion mittels Quota-Verfahren**

Bei einer angenommenen Populationsgröße von 41.250 Personen, einem Konfidenzniveau von 95 Prozent und einer Fehlertoleranz von 5 Prozent, sollte eine Zufallsstichprobe 381 Personen umfassen.

Zufallsgesteuerte Auswahlverfahren von Probanden für die Stichprobe schließen sich methodisch aufgrund der (1) Unsicherheit über die aktuelle Grundgesamtheit sowie aufgrund des (2) fehlenden Zugangs zu einer Kontakt-Datenbank wie zimpel oder der des DJV aus. Da allerdings Daten zur Verteilung soziodemografischer Merkmale innerhalb der Grundgesamtheit gegeben sind (Tab. 2) wurde zur Erhebung das Quota-Verfahren gewählt.

Die Stichprobenkonstruktion mittels Quoten-Auswahl stellt eine geschichtete willkürliche Auswahl dar. Um mittels Quotierung die Repräsentativität einer Stichprobe zu sichern, werden drei Qualitätskriterien überprüft: Erstens die Bekanntheit der Quotierungsmerkmale in der Grundgesamtheit, zweitens eine hohe Korrelation zwischen Quotierungsmerkmalen und Untersuchungsmerkmalen, drittens die leichte Erfassbarkeit der Quotierungsmerkmale. (Kromrey et al. 2016, S. 271–277)

Als Quotierungsmerkmale wurden Geschlecht, Alter und Anstellungsverhältnis aufgrund der größten Vergleichbarkeit mit bestehenden Studien (Tab. 2) festgelegt. Die Verteilung der Merkmale in der Grundgesamtheit wurde übernommen von Hanitzsch und Rick (2021), welche von Oktober bis

Dezember 2020 über den Verteiler des Deutschen Journalisten-Verbandes 1.055 Journalistinnen und Journalisten befragt haben.

Tab. 3: Quotierungsplan nach Geschlecht, Alter und Anstellungsverhältnis

| Quotierungsmerkmal | Quote (Ziel-n = 381) | Verteilung in GG | Ausschöpfung |
| --- | --- | --- | --- |
| Geschlecht | 235 männliche Personen | 61,7 % | 173 (73,62 %) |
| Geschlecht | 144 weibliche Personen | 37,7 % | 153 (106,25 %) |
| Alter | 32 Personen (bis 29 Jahre) | 8,4 % | 87 (271,88 %) |
| Alter | 75 Personen (30 – 39 Jahre) | 19,8 % | 85 (113,33 %) |
| Alter | 64 Personen (40 – 49 Jahre) | 16,8 % | 68 (106,25 %) |
| Alter | 210 Personen über 50 Jahre | 55 % | 80 (38,10 %) |
| Anstellungsverhältnis | 233 freie Journalisten | 61,1 % | 212 (90,99 %) |
| Anstellungsverhältnis | 148 angestellte Journalisten | 38,9 % | 117 (79,05 %) |

Bei Betrachtung der Quotenausschöpfung zeigt sich eine Überrepräsentanz von Personen unter 29 Jahren (271,88 Prozent) sowie eine deutliche Unterrepräsentanz von nur 38 Prozent in der Altersquote der über 55-Jährigen. Um diese Datenverzerrung auszugleichen, wird eine Fallgewichtung vorgenommen. Für die getroffenen Quotierungsmerkmale Alter, Anstellungsverältnis und Geschlecht wurden Gewichtungsvariablen berechnet und als einfaches Multiplikat ohne Anpassungsgewichte fallweise angewendet (siehe Anhang). Die Gründe für diese Verzerrung sowie die resultierenden methodischen Implikationen werden in Kapitel 5 diskutiert.

Fehlende 0,6 Prozent (3 Personen) bei der Verteilung des Geschlechts in der Grundgesamtheit sind auf das diverse Geschlecht zurückzuführen. Aufgrund der Unterrepräsentanz wird das diverse Geschlecht für die statistische Analyse ausgeschlossen.

**Akquise von Probanden und Datenerhebung durch ein Web-Survey**

Zur Akquise von Probanden, die hauptberuflich journalistisch tätig sind, wurde eine Kontakt-Datenbank aus 5.705 Personen aufgebaut. Zum Aufbau der Kontaktdatenbank wurden die Autoren-, Kontakt und Impressumsseiten der 100 reichweitenstärksten Nachrichten-Webseiten (IVW) sowie Portale für Medienschaffende durchsucht (kress: 2605, Portale: 1587, torial: 1513).

Vom 02.12.2020 bis zum 03.12.2020 wurden an diese 5.705 Personen in mehreren Paketen – mit Blick auf automatisierte Spam-Erkennungssysteme – E-Mails mit dem Umfrage-Link verschickt. Eine Erinnerung wurde am 16.12. mit anderer E-Mail-Adresse verschickt. Bis zum Ende der Umfrage am 31. Dezember 2020 haben 524 einzigartige Personen den Umfrage-Link angeklickt. 432 E-Mail-Adressen waren ungültig, 93 Personen haben sich aus dem Verteiler abgemeldet. Vollständig abgeschlossen haben den Fragebogen 385 Personen. Das bereinigte Gesamtsample beträgt damit nach Abzug stichprobenneutraler und systematischer Ausfälle 5.180 Elemente. Die Ausschöpfungsquote beträgt 7,4 Prozent.

Als Datenerhebungsmethode wurde ein sogenannter Web-Survey, also eine internetgestützte Befragungsmethode, gewählt. Bei der formalen Gestaltung des Fragebogens wurden die Empfehlungen der Methoden-Literatur (Schnell et al. 2018, S. 374–375) umgesetzt.

**Fragebogenaufbau und Messinstrumente**

Nach Einleitung und Beschreibung der Studie wurde durch zwei Ausschlussfragen die Passung des potenziellen Umfrageteilnehmers geprüft. Auf die dichotome Frage „Waren Sie im Jahr 2020 journalistisch tätig?" folgte bei Verneinung zum einen die Frage nach der eigenen Berufsbezeichnung (Freitext) und zum anderen die dichotome Frage, ob die jeweilige Person im Jahr 2020 in einer Redaktion gearbeitet hat. Aus Konsistenzgründen wurde bei Zustimmung zu diesem Fragebogenteil auch die Redaktionsgröße (Merkmalsausprägungen: 0-5, 6-15, 16-30, 31-50, 51-100, 101-200, mehr als 200) abgefragt. Die Fragen werden dem Konstrukt 1A (Journalistische Tätigkeit) zugeordnet.

Für das Konstrukt 2A (Plattformbereitschaft) wurde als Messinstrument eine fünfstufige Likert-Skala mit den Ausprägungen „trifft zu, trifft eher zu, teils-teils, trifft eher nicht zu, trifft nicht zu" gewählt. Die Frage nach der nebenberuflichen Nutzungsbereitschaft einer journalistischen Plattform wurde nur gezeigt sofern die Frage nach der hauptberuflichen Nutzungsbereitschaft mit „trifft eher nicht zu", „trifft nicht zu" oder „teils-teils" beantwortet wurde.

Im folgenden Fragebogenteil wurde mit einer fünfstufigen Attraktivitätsskala (völlig unattraktiv, unattraktiv, neutral, attraktiv, ideal) die wahrgenommene Attraktivität von vier Plattform-Finanzierungsmodellen (Konstrukt 2B) ermittelt. Um möglichen Theorie-Paradigmatismus zu vermeiden schließt der Fragebogenteil mit einer Freitext-Frage zu alternativen Finanzierungsmodellen.

Das Konstrukt 3A (Normative Anforderungen) wurde mit einer fünfstufige Wichtigkeitsskala nach Likert in den Ausprägungen „überhaupt nicht wichtig, etwas wichtig, relativ wichtig, sehr wichtig, äußerst wichtig" gemessen.

Um die inhaltlich konträren Anforderungen an eine Journalismusplattform (Konstrukt 3B) zu erheben wurde das Semantische Differential als Messinstrument verwendet.

Die Nutzungshäufigkeit und Nutzungsintention technologischer Innovationen im Journalismus wurde in den Konstrukten 4A (Multimedia), 4B (KI), 4C (Feedback) in abwechselnder Reihenfolge der Frage nach Häufigkeit und Intention mit einer fünfstufigen Häufigkeitsskala sowie einer klassischen Likert-Skala gemessen.

Für den zweiten Teil des Konstrukts „Journalistische Tätigkeit" wurden für die Fragen nach veröffentlichten Medienformen, Arbeitsverhältnis, der wöchentlichen Arbeitszeit sowie der Anzahl an Veröffentlichungen aus Vergleichbarkeitsgründen die Skalen von Gautschi (2016) übernommen.

Zum Abschluss des Fragebogens wurden Soziodemografika (Konstrukt 5) und entsprechend die Quotierungsmerkmale abgefragt. Um die höchstmögliche Vergleichbarkeit zu historischen Journalistenbefragungen herstellen zu können, wurden die Messinstrumente zu Einkommen und Bildung von Weischenberg et al. (2006) übernommen.

# 4 Ergebnisse

Journalistinnen und Journalisten haben eine Bereitschaft auf einer Journalismusplattform hauptberuflich zu publizieren. Freie haben eine höhere Bereitschaft als Angestellte. Ein Plattform-Abonnement wird attraktiver bewertet als die Werbefinanzierung einer Plattform. Angestellte Journalisten finden hingegen eine Werbefinanzierung attraktiver als Freie. Es ist Journalistinnen und Journalisten wichtig, dass die Plattform aus Europa oder Deutschland stammt und besonderen Wert auf Datenschutz legt. Multimediale Darstellungsformen und interaktive Elemente werden gelegentlich, oft oder immer genutzt. Aneinandergereihte Bilder und Videos (Storys oder Reels) werden überwiegend nicht genutzt. KI-Software als Produktionshilfe sowie Analyse-Tools zur Auswertung eigener Veröffentlichungen werden selten oder nie genutzt. Mit Ausnahme von Storys oder Reels intendieren Journalistinnen und Journalisten multimediale Darstellungsformen und interaktive Elemente in Zukunft häufiger zu nutzen. Gegenüber Software zur Dokumentation und Darstellung des Rechercheprozesses sowie zur Analyse von Kennzahlen eigener Veröffentlichungen sind sie aufgeschlossen. Software als Unterstützung zur Textproduktion, Bildauswahl oder für Überschriftenvorschläge wird mehrheitlich abgelehnt.

Tab. 4: Ergebnisse der Hypothesentests[2]

| Plattform-Bereitschaft | Journalistinnen und Journalisten sind bereit auf einer Journalismusplattform hauptberuflich zu publizieren. | angenommen |
|---|---|---|
| | Berufseinsteiger haben eine höhere Bereitschaft. | verworfen |
| | Freie haben eine höhere Bereitschaft als Angestellte. | angenommen |
| Plattform-Finanzierung | Ein Plattform-Abonnement wird attraktiver bewertet als eine Werbefinanzierung. | angenommen |
| | Eine Klubfinanzierung wird attraktiver bewertet als ein Abonnement. | abgelehnt |
| | Mirco-Payments werden als völlig unattraktiv, unattraktiv oder neutral bewertet. | angenommen |
| | Freie finden eine Klubfinanzierung attraktiver als Angestellte | verworfen |
| | Freie finden Micro Payments attraktiver als Angestellte. | verworfen |
| | Angestellte finden ein monatliches Abonnement attraktiver als Freie. | verworfen |
| | Angestellte Journalisten finden eine Werbefinanzierung attraktiver als Freie. | angenommen |

---

[2] Alle Hypothesentests wurden auch ohne Fallgewichtung durchgeführt. Es konnten keine inhaltlichen Aussagenunterschiede festgestellt werden.

| | | |
|---|---|---|
| Anforderungen | Es ist überhaupt nicht wichtig, etwas wichtig oder relativ wichtig, dass die Plattform besonderen Wert auf Datenschutz legt. | abgelehnt |
| | Es ist überhaupt nicht wichtig, etwas wichtig oder relativ wichtig, dass die Plattform unabhängig von etablierten Tech-Giganten (wie Google und Facebook) ist. | abgelehnt |
| | Es ist überhaupt nicht wichtig, etwas wichtig oder relativ wichtig, dass die Plattform öffentlich-rechtlich ist. | angenommen |
| | Es ist überhaupt nicht wichtig, etwas wichtig oder relativ wichtig, dass die Plattform von den etablierten Verlagen entwickelt wird. | angenommen |
| Nutzungshäufigkeit Multimedia | Die Mehrheit nutzt aneinandergereihte Bilder und Videos (sog. Story oder Reels) gelegentlich, oft oder immer. | verworfen |
| | Die Mehrheit nutzt interaktive Elemente gelegentlich, oft oder immer. | angenommen |
| | Die Mehrheit nutzt multimediale Darstellungsformen gelegentlich, oft oder immer. | angenommen |
| Nutzungsintention Multimedia | Die Mehrheit intendiert aneinandergereihte Bilder und Videos in Zukunft nicht häufiger zu nutzen. | abgelehnt |
| | Die Mehrheit intendiert interaktive Elemente in Zukunft nicht häufiger zu nutzen. | abgelehnt |
| | Die Mehrheit intendiert eine multimediale Darstellung in Zukunft nicht häufiger zu nutzen. | abgelehnt |
| Nutzungshäufigkeit Software | Die Mehrheit nutzt gelegentlich, oft oder immer KI-Software als Produktionshilfe für ihre journalistische Arbeit. | abgelehnt |
| | Die Mehrheit dokumentiert ihren Rechercheprozess nie oder selten ausführlich. | abgelehnt |
| | Die Mehrheit nutzt nie oder selten Analyse-Tools, um Kennzahlen ihrer Veröffentlichungen auszuwerten. | abgelehnt |
| Nutzungsintention Software | Die Mehrheit würde Software als Produktionshilfe in Zukunft nutzen, insofern die Qualität besser würde. | abgelehnt |
| | Die Mehrheit würde in Zukunft umfangreiche Angaben zum Rechercheprozess machen, wenn eine Plattform dies vereinfachen würde. | angenommen |
| | Die Mehrheit würde ein speziell zugeschnittenes Analyse-Dashboard nutzen. | angenommen |

## 4.1 Berufliche Nutzungsbereitschaft einer Journalismusplattform

*$H_0$: Journalistinnen und Journalisten sind nicht bereit auf einer Plattform hauptberuflich zu publizieren.*

*$H_1$: Journalistinnen und Journalisten sind bereit auf einer Plattform hauptberuflich zu publizieren.*

66,6 Prozent der Journalistinnen und Journalisten geben an auf einer Plattform hauptberuflich publizieren zu wollen, insofern diese den eigenen Lebensunterhalt sichert. 17,6 Prozent können sich dies nicht oder gar nicht vorstellen. 16,2 Prozent sind unentschlossen. Der Mittelwert der Antworten liegt bei 3,82 bei einer Standardabweichung von 1,267. Ein t-Test ergibt, dass die Differenz zum mittleren Skalenwert 3 („teils-teils") signifikant ist ($t(294) = 11{,}082$, $p < 0{,}001$). Damit kann *$H_1$* vorläufig angenommen werden: Die Bereitschaft unter Journalistinnen und Journalisten auf einer Plattform hauptberuflich zu veröffentlichen, insofern diese ihren Lebensunterhalt sichert, ist gegeben.

*$H_0$: Es gibt keinen Unterschied bei der hauptberuflichen Nutzungsbereitschaft einer Journalismusplattform zwischen Berufseinsteigern und Berufserfahrenen.*

*$H_1$: Berufseinsteiger haben eine höhere hauptberufliche Nutzungsbereitschaft für eine Journalismusplattform als Berufserfahrene.*

Als Berufseinsteiger gelten im Journalismus nach Weischenberg et al. (2006) Personen unter 36 Jahren. 69 Befragte (22,4 %) gelten damit als Berufseinsteiger. 73,2 Prozent der Berufseinsteiger und 64,1 Prozent der berufserfahrenen Journalistinnen und Journalisten haben eine hohe oder sehr hohe Plattformnutzungsbereitschaft.

Mit dem t-Test für zwei unabhängigen Stichproben kann getestet werden, ob bei einer Veränderung der unabhängigen Variable, die zwei Ausprägungen bzw. Gruppen hat, signifikante Änderungen bei der abhängigen Variable auftreten – ob also ein Gruppenunterschied vorliegt. Um einen T-Test für zwei unabhängigen Stichproben durchführen zu können müssen (1) die beiden Gruppen bzw. Stichproben unabhängig sein, (2) die abhängige Variable metrisch skaliert sein, (3) die Variablen normalverteilt sein und (4) die Varianz innerhalb der Gruppen ähnlich sein.

Die Varianz der abhängigen Variable ist in den Gruppen nicht gleich. Die abhängige Variable der hauptberuflichen Nutzungsbereitschaft wurde mit einer Likert-Skala gemessen. Somit werden zwei der vier Voraussetzungen für einen T-Test für zwei unabhängige Stichproben verletzt, weshalb der Mann-Whitney-U-Test zum Testen der Hypothesen verwendet wird.

Da n > 30 wird die 2-seitige asymptotische Signifikanz berichtet. Da P = 0,097 > 0,05 wird $H_0$ beibehalten. Es gibt keinen signifikanten Unterschied bei der hauptberuflichen Nutzungsbereitschaft einer Journalismusplattform zwischen Berufseinsteigern und Berufserfahrenen.

*$H_0$: Es gibt zwischen Freien und Angestellten keinen Unterschied bei der hauptberuflichen Nutzungsbereitschaft für eine Journalismusplattform.*

*$H_1$: Freie haben eine höhere hauptberufliche Nutzungsbereitschaft für eine Journalismusplattform als Angestellte.*

Als Freie gelten alle selbständigen und freiberuflichen Journalistinnen und Journalisten sowie Pauschalisten und Feste Freie ohne ein hauptberufliches, versicherungspflichtiges Beschäftigungsverhältnis bei einer Redaktion. 211 Befragte (68,4 %) gelten demnach als freie Journalisten, 97 (31,6 %) als Angestellte. 54,5 Prozent der angestellten Journalistinnen und Journalisten haben eine hohe oder sehr hohe Plattformnutzungsbereitschaft, wohingegen die Bereitschaft in der Gruppe der Freien mit 71,6 Prozent deutlich höher ist.

Zum Test auf den Unterschied der zentralen Tendenz innerhalb der beiden Gruppen wurden zunächst die Voraussetzungen für einen t-Test für zwei unabhängigen Stichproben überprüft. Die Gruppen sind unabhängig voneinander, die abhängige Variable likert-skaliert, die Variablen normalverteilt und die Varianz innerhalb der Gruppen ähnlich. Damit wurde eine Voraussetzung, die metrische Skalierung der abhängigen Variable, verletzt, weshalb der Mann-Whitney-U-Test verwendet wird.

Da P = 0,006 < 0,05 wird $H_1$ vorläufig angenommen: Freie Journalistinnen und Journalisten haben eine signifikant höhere hauptberufliche Nutzungsbereitschaft für eine Journalismusplattform als angestellte Journalistinnen und Journalisten.

### 4.2 Attraktivität von Finanzierungsmodellen einer Journalismusplattform

Folgend wird die wahrgenommene Attraktivität von Finanzierungsmodellen für eine Journalismusplattform dargestellt. Erhoben wurde die Einstellung zu Micro Payments, zu Werbung, zu einer Klubfinanzierung sowie zu monatlichen Plattform-Abonnements. Bei einem Plattform-Abonnement können Nutzerinnen und Nutzer eine Plattform mit einem monatlichen Beitrag abonnieren, sodass sie Zugriff auf alle Inhalte der Plattform erhalten. Beim Modell der Klubfinanzierung abonnieren Nutzerinnen und Nutzer monatlich nur eine Journalistin oder einen Journalisten. Im durch Werbung finanzierten Plattform-Modell werden produzierte oder eingekaufte journalistische Inhalte über Werbung finanziert. Micro Payments zeichnen sich dadurch aus, dass Rezipierende pro angeklickten Beitrag einen kleinen Betrag bezahlen.

**Wie bewerten Journalistinnen und Journalisten Plattformfinanzierungsmodelle?**

Die Mehrheit der Journalistinnen und Journalisten sieht das monatliche Abonnement einer Plattform als attraktivstes Finanzierungsmodell (MW = 4,06), gefolgt von der Klubfinanzierung (MW = 3,73). Ein Werbeerlösmodell wird im Vergleich 20 Prozent weniger attraktiv bewertet (MW = 3,1). Ähnlich zur Werbung werden im Mittel (MW = 2,97) Zahlungen für einzelne Veröffentlichungen (sog. Micro-Payments) als neutral gesehen, wobei die Antworten eine bimodale Verteilung aufweisen. 40,4 Prozent der Journalistinnen und Journalisten bewerten Micro Payments als unattraktiv oder völlig unattraktiv, wohingegen 44,8 Prozent dieses Modell als attraktiv oder ideal bewerten.

Abb. 1: Boxplots zu den Antworten auf die Plattformfinanzierungsmodelle: Klubfinanzierung, Micro Payments, Abonnement und Werbung

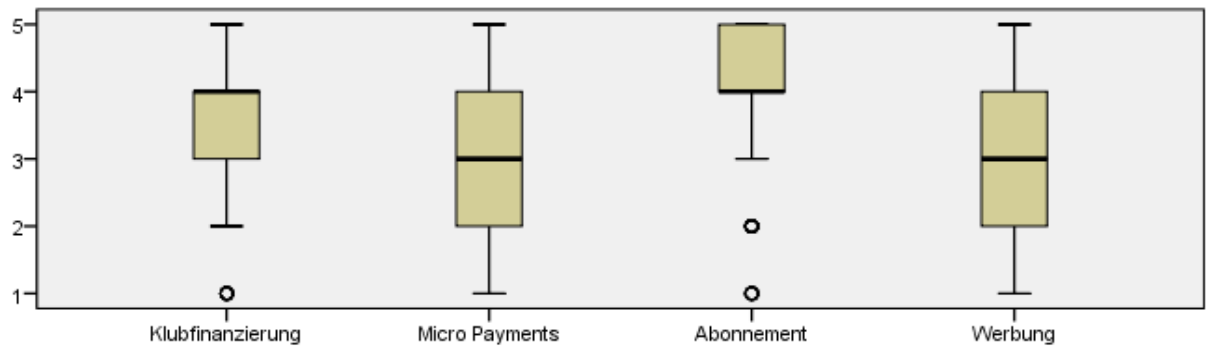

*H$_0$: Zwischen den Finanzierungsmodellen Abonnement und Werbung gibt es keinen Unterschied.*

*H$_1$: Ein Abonnement als Finanzierungsmodell einer journalistischen Plattform wird von Journalist:innen attraktiver bewertet als eine Werbefinanzierung.*

Zum Test auf den Unterschied der zentralen Tendenz innerhalb der beiden Gruppen wurden zunächst die Voraussetzungen für einen t-Test für zwei abhängige Gruppen überprüft. Es liegen zwei abhängige Gruppen bzw. Stichproben vor, die Variablen sind likert-skaliert und damit nur quasi-metrisch, die Differenzen der gepaarten Werte sind normalverteilt. Damit wird eine Voraussetzung, die metrische Skalierung Variablen, verletzt, weshalb der Wilcoxon-Test verwendet wird.

Da P = 0,000 < 0,001 wird H1 vorläufig angenommen: Es gibt einen signifikanten Unterschied in der Bewertung der Attraktivität eines Abonnements als Finanzierungsmodell gegenüber einer Werbefinanzierung. Bei Z = 8,621 und N = 297 ist r = 0,50, was einen starken Effekt bedeutet.

*H$_0$: Zwischen Klubfinanzierung und Abonnement gibt es keinen Unterschied.*

*H$_1$: Eine Klubfinanzierung als Finanzierungsmodell einer journalistischen Plattform wird von attraktiver bewertet als ein Abonnement.*

Wie im vorangegangen Hypothesentest werden die Bedingungen eines t-Test für zwei abhängige Gruppen verletzt, weshalb der Wilcoxon-Test verwendet wird. Da die mittlere Differenz -0,33 beträgt und P = 0,000 < 0,001 wird H$_1$ vorläufig abgelehnt: Ein Abonnement einer Plattform wird signifikant attraktiver bewertet als eine Klubfinanzierung. Bei Z = 4,940 und N = 295 ist r = 0,29, was einen mittleren Effekt bedeutet.

*H$_0$: Mirco-Payments werden als attraktiv oder ideal bewertet.*

*H$_1$: Mirco-Payments werden als völlig unattraktiv, unattraktiv oder neutral bewertet.*

Wie im vorangegangen Hypothesentest werden die Bedingungen eines t-Test für zwei abhängige Gruppen verletzt, weshalb der Wilcoxon-Test verwendet wird. Da die mittlere Differenz -1,034 beträgt und p < 0,001 wird $H_1$ vorläufig abgelehnt.

**Bewerten Freie die Plattformfinanzierung anders als angestellte Journalisten?**

Die Autoren dieser Studie stellen die Hypothese auf, dass es einen Unterschied in der Bewertung der Attraktivität von Finanzierungsmodellen zwischen freien und angestellten Journalistinnen und Journalisten gibt (siehe Tab. 5). Wie im vorangegangen Hypothesentest werden die Bedingungen eines t-Test für zwei abhängige Gruppen verletzt, weshalb der Wilcoxon-Test verwendet wird.

Tab. 5: Hypothesentests zu Gruppenunterschieden zwischen Freien und Angestellten bei der Bewertung von Finanzierungsmodellen

| $H_0$ | Es gibt keinen Gruppenunterschied. | |
|---|---|---|
| $H_1a$ | Freie Journalisten finden eine Klubfinanzierung als Finanzierungsmodell einer journalistischen Plattform attraktiver als angestellte Journalisten. | P = 0,909 > 0,05, damit nicht signifikant und verworfen |
| $H_1b$ | Freie Journalisten finden Micro Payments als Finanzierungsmodell einer journalistischen Plattform attraktiver als angestellte Journalisten. | P = 0,136 > 0,05, damit nicht signifikant und verworfen |
| $H_1c$ | Angestellte Journalisten finden ein monatliches Abonnement als Finanzierungsmodell einer journalistischen Plattform attraktiver als freie Journalisten. | P = 0,138 > 0,05, damit nicht signifikant und verworfen |
| $H_1d$ | Angestellte Journalisten finden eine Werbefinanzierung als Finanzierungsmodell einer journalistischen Plattform attraktiver als freie Journalisten. | P = 0,001 < 0,05, damit signifikant und vorläufig angenommen |

Die Hypothesen $H_1a$ - $H_1c$ werden aufgrund mangelnder Signifikanz verworfen. Ein Gruppenunterschied im Mittelwert der Attraktivitätsbewertung lässt sich lediglich bei der Werbefinanzierung bestimmen: Angestellte Journalistinnen und Journalisten bewerten die Werbefinanzierung einer journalistischen Plattform signifikant attraktiver als freie Journalisten.

In einer abschließenden offenen Frage nach weiteren attraktiven Finanzierungsmodellen wurden 77 gültige Texteingaben getätigt. Die jeweiligen Texteingaben wurden paraphrasiert und induktiv kodiert. Aus der Kodierung haben sich zwölf Kategorien für Finanzierungsmodelle ergeben. Werbemodell, Abonnement, Klubfinanzierung sowie Micro Payments wurden trotz vorheriger Abfrage dieser Modelle 20-mal genannt, meist spezifiziert. 18 Nennungen entfallen auf öffentlich-rechtlichen Journalismus (fünf davon: Staatliche Finanzierung). Eine Finanzierung über Spenden, die so genannte Solidarfinanzierung, wurde zwölf Mal genannt. Stiftungsmodelle können sich 13 Befragte vorstellen. Zehn Befragte explizieren Crowdfunding als attraktives Plattformfinanzierungsmodell, sieben Sponsoring und drei ein Genossenschaftsmodell. Vier Befragte beschreiben ein Modell bei dem Nutzerinnen und Nutzer mehrere Themen gebündelt zu einem festen Monatspreis abonnieren können.

## 4.3 Normative Anforderungen an eine Journalismusplattform

Jeder vierten Journalistin, jedem vierten Journalisten (24,7 Prozent) ist es überhaupt nicht wichtig oder nur etwas wichtig, dass eine Journalismusplattform in Europa oder Deutschland entwickelt wird. Die Hälfte der Journalistinnen und Journalisten (49,5 Prozent) empfindet diesen Aspekt als sehr oder äußerst wichtig. Bei einem Mittelwert von 3,37 (SD = 1,278) und einem Median von 3 lässt sich sagen, dass die Mehrheit der Journalistinnen und Journalisten Europa oder Deutschland als Herkunft für eine journalistische Plattform als relativ wichtig bewertet.

*$H_0$: Es ist sehr wichtig oder äußerst wichtig, dass die Plattform aus Europa oder Deutschland ist.*

*$H_1$: Es ist überhaupt nicht wichtig, etwas wichtig oder relativ wichtig dass die Plattform aus Europa oder Deutschland ist.*

Zum Test der Hypothese H1 wurde ein T-Test bei einer Stichprobe durchgeführt. Bei einem Trennwert von 3 ergibt sich eine mittlere Differenz von 0,367. $H_0$ wird damit beibehalten (t(297) = 4,951, p < ,001).

11,7 Prozent der Befragten sehen den Datenschutz einer Journalismusplattform als überhaupt nicht wichtig oder etwas wichtig. Drei von vier Journalistinnen und Journalisten (71,4 Prozent) empfinden Datenschutz als sehr wichtig oder äußerst wichtig. Der Mittelwert liegt bei 3,95 (SD = 1,032), was sich in die Skalenausprägung „sehr wichtig" übersetzen lässt.

*$H_0$: Es ist sehr wichtig oder äußerst wichtig, dass die Plattform besonderen Wert auf Datenschutz legt.*

*$H_1$: Es ist überhaupt nicht wichtig, etwas wichtig oder relativ wichtig, dass die Plattform besonderen Wert auf Datenschutz legt.*

Zum Test der Hypothese $H_1$ wurde ein T-Test bei einer Stichprobe durchgeführt. Bei einem Trennwert von 3 ergibt sich eine mittlere Differenz von 0,950. $H_0$ wird damit beibehalten (t(302) = 16,029, p < ,001).

Fast 85 Prozent der Journalistinnen und Journalisten ist es ist relativ wichtig, sehr wichtig oder äußerst wichtig, dass eine Journalismusplattform unabhängig von sog. Tech-Giganten wie Google oder Facebook entwickelt wird. Bei einem Mittelwert von 3,91 (SD = 1,302) und einem Median von 4 ist diese Unabhängigkeit der Mehrheit sehr wichtig.

*$H_0$: Es ist sehr wichtig oder äußerst wichtig, dass die Plattform unabhängig von etablierten Tech-Giganten (wie Google und Facebook) ist.*

*$H_1$: Es ist überhaupt nicht wichtig, etwas wichtig oder relativ wichtig, dass die Plattform unabhängig von etablierten Tech-Giganten (wie Google und Facebook) ist.*

Zum Test der Hypothese $H_1$ wurde ein T-Test bei einer Stichprobe durchgeführt. Bei einem Trennwert von 3 ergibt sich eine mittlere Differenz von 0,913. $H_0$ wird damit beibehalten (t(298) = 12,116, p < ,001): Der Mehrheit ist die Unabhängigkeit von Tech-Giganten signifikant wichtig.

Die Option, dass eine Journalismusplattform im oder durch das öffentlich-rechtliche Rundfunksystem entwickelt wird, lehnen rund 35 Prozent der Journalistinnen und Journalisten vollständig ab. 26,5 Prozent sehen eine „öffentlich-rechtliche Plattform" als sehr wichtig oder äußerst wichtig an. Bei einem Mittelwert von 2,54 (SD = 1,44) und einem Median von 2 sieht die Mehrheit die Option einer öffentlich-rechtlichen Plattform als „etwas wichtig" an.

*$H_0$: Es ist sehr wichtig oder äußerst wichtig, dass die Plattform öffentlich-rechtlich ist.*

*$H_1$: Es ist überhaupt nicht wichtig, etwas wichtig oder relativ wichtig, dass die Plattform öffentlich-rechtlich ist.*

Es wurde ein T-Test bei einer Stichprobe durchgeführt. Bei einem Trennwert von 3 ergibt sich eine mittlere Differenz von -0,462. $H_1$ wird vorläufig angenommen (t(286) = -5,439, p < ,001).

*$H_0$: Es ist sehr oder äußerst wichtig, dass die Plattform von den etablierten Verlagen entwickelt wird.*

*$H_1$: Es ist überhaupt nicht wichtig, etwas wichtig oder relativ wichtig, dass die Plattform von den etablierten Verlagen entwickelt wird.*

Bei einem Mittelwert von 2,37 (SD = 1,26) und einem Median von 2 sieht die Mehrheit der Journalistinnen und Journalisten eine Plattform-Entwicklung durch etablierte Verlage als etwas wichtig an. Äußert wichtig oder sehr wichtig bewerten 19,4 Prozent eine solche Verlagsplattform.

Zum Test der Hypothese $H_1$ wurde ein Einstichproben-T-Test durchgeführt. Es ergibt sich eine mittlere Differenz von -0,660. $H_1$ wird vorläufig angenommen (t(298) = -9,065, p < ,001).

Neben den Rahmenbedingungen einer Plattform sollen Veröffentlichungen der Journalisten überwiegend mit ihrem eigenen Namen statt mit einer Redaktionsmarke dargestellt werden. Ein „One-Stop-Shop" für journalistische Inhalte – also die Möglichkeit alle Nachrichten an einem Ort lesen zu können – wird gegenüber mehrerer Insellösungen bevorzugt. Die transparente Darstellung der Auswahlentscheidungen von Empfehlungsalgorithmen sehen Journalisten als klaren Vorteil für den Leser. Dagegen wird eine algorithmische Qualitätsbewertung einzelner Veröffentlichungen als Eingriff in die journalistische Freiheit betrachtet. Bei den Fragen nach einer personalisierten Startseite und einer generellen Vorauswahl von „Qualitätsmedien" divergieren die Meinungen der Journalisten stark. (siehe Abb. 2)

Abb. 2: Polaritätsprofil mittels Boxplots für Anforderungen an eine Journalismusplattform

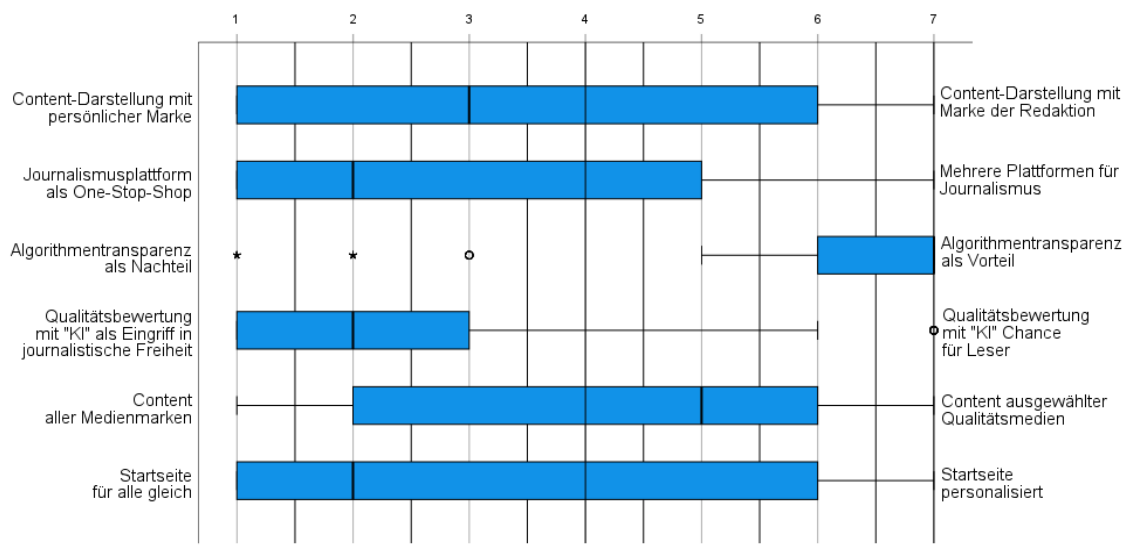

### 4.4 Nutzungshäufigkeit und -intention multimedialer und interaktiver Darstellungsformen

Aneinandergereihte Bilder und Videos werden von 44,5 Prozent der Journalistinnen und Journalisten nie genutzt, 26,7 Prozent nutzen diese Darstellungsform selten, 15,9 Prozent gelegentlich. 12,9 Prozent nutzen die sogenannten Storys bzw. Reels oft oder immer.

*$H_0$: Die Mehrheit nutzt aneinandergereihte Bilder und Videos nie oder selten.*

*$H_1$: Die Mehrheit nutzt aneinandergereihte Bilder und Videos gelegentlich, oft oder immer.*

Ein t-Test bei einer Stichprobe mit dem Trennwert 2 ergibt eine mittlere Differenz von -0,13. $H_0$ wird aufgrund fehlender Signifikanz beibehalten (t(307) = -0,204, p > ,05).

Interaktive Elemente werden von Journalistinnen und Journalisten im Mittel (MW = 2,33; SD = 1,022) selten genutzt. 58,1 Prozent nutzen interaktive Inhalte in ihren Veröffentlichungen nie oder nur selten, 28,6 Prozent gelegentlich, 13,3 Prozent oft oder immer.

*$H_0$: Die Mehrheit nutzt interaktive Elemente nie oder selten.*

*$H_1$: Die Mehrheit nutzt interaktive Elemente gelegentlich, oft oder immer.*

Zum Test der Hypothese $H_1$ wurde ein Einstichproben-T-Test durchgeführt. Es ergibt sich eine mittlere Differenz von 0,333. $H_0$ wird vorläufig angenommen (t(307) = 5,728, p < ,001).

Multimediale Darstellungsformen nutzen Journalistinnen und Journalisten im Mittel gelegentlich, was ein Mittelwert von 2,7 (SD = 1,189) und ein Median von 3 zeigen. 52,2 Prozent nutzen Multimedia gelegentlich, oft oder immer, wohingegen 47,8 Prozent nie oder selten mehr als eine Medienform in ihren Veröffentlichungen nutzen.

*$H_0$: Die Mehrheit nutzt multimediale Darstellungsformen nie oder selten.*

*H₁: Die Mehrheit nutzt multimediale Darstellungsformen gelegentlich, oft oder immer.*

Ein t-Test bei einer Stichprobe mit dem Trennwert 2 ergibt eine mittlere Differenz von 0,696. H₀ wird beibehalten (t(307) = 10,274, p < ,001).

Aus den drei Items zur Nutzungshäufigkeit multimedialer Darstellungsformen wurde eine Skala gebildet. Als Reliabilitätsmaß zur Bestimmung der internen Konsistenz wurde Cronbachs Alpha gewählt. Für diese Skala ist Cronbachs α = 0,654 und weist damit eine gerade noch zufriedenstellende interne Konsistenz (0,7 > α > 0,6) auf.

**Nutzungsintention multimedialer und interaktiver Darstellungsformen**

Zwei Drittel der Journalistinnen und Journalisten möchte in zukünftigen journalistischen Veröffentlichungen eine multimediale Darstellungsform (71,7 Prozent) und interaktive Elemente (67,8 Prozent) häufiger nutzen. Sogenannte Storys oder Reels – aneinandergereihte Bilder und Videos – intendieren nur 37,3 Prozent in Zukunft häufiger zu nutzen. Eine ablehnende Haltung zu multimedialen und interaktiven Darstellungsformen nimmt nur jede siebte Journalistin bzw. jeder siebte Journalist ein. Im Mittel wird die Nutzungsintention multimedialer Darstellungsformen mit 3,92 Punkten bewertet (SD = 1,113) bei einer Skala von trifft nicht zu (1) bis trifft zu (5). Die Nutzungsintention interaktiver Elemente erreicht einen Wert von 3,78 (SD = 1,165) und die Nutzungsintention von Storys 2,97 (SD = 1,323), womit dieser Wert innerhalb des Konstruktes der niedrigste ist. 40 Prozent der Journalistinnen und Journalisten können sich nicht vorstellen in Zukunft Storys oder Reels für ihre journalistische Arbeit zu nutzen.

Tab. 6: Hypothesentests zur Nutzungsintention multimedialer und interaktiver Darstellungsformen

| H₀ | Die Mehrheit würde die abgefragte Darstellungsform in Zukunft häufiger nutzen. | |
|---|---|---|
| H₁ₐ | Die Mehrheit intendiert aneinandergereihte Bilder und Videos in Zukunft nicht häufiger zu nutzen. | t(307) = -13,641, p < ,001 H₀ beibehalten |
| H₁ᵦ | Die Mehrheit intendiert interaktive Elemente in Zukunft nicht häufiger zu nutzen. | t(307) = -3,279, p < ,05 H₀ beibehalten |
| H₁꜀ | Die Mehrheit intendiert eine multimediale Darstellung in Zukunft nicht häufiger zu nutzen. | t(306) = -1,266, p > ,05 H1 verworfen |

Aus den drei Items zur Nutzungsintention multimedialer Darstellungsformen wurde eine Skala gebildet. Als Reliabilitätsmaß zur Bestimmung der internen Konsistenz wurde Cronbachs Alpha gewählt. Für diese Skala ist Cronbachs α = 0,667 und weist damit eine gerade noch zufriedenstellende interne Konsistenz (0,7 > α > 0,6) auf.

## 4.5 Nutzungshäufigkeit und -intention von Software im Produktionsprozess

KI-Software als Produktionshilfe können z. B. Überschriftenvorschläge, Textbausteine, eine automatische Bildauswahl oder die Optimierung von Text sein. 60,1 Prozent der Journalistinnen und Journalisten nutzen solche Software nie, 19,2 Prozent selten. Bei einem Mittelwert von 1,78 (SD = 1,184) und einem Median von 1 nutzt die überwiegende Mehrheit der Journalistinnen und Journalisten nie KI-Software als Produktionshilfe.

*$H_0$: Die Mehrheit nutzt nie oder selten KI-Software als Produktionshilfe.*

*$H_1$: Die Mehrheit nutzt gelegentlich, oft oder immer KI-Software als Produktionshilfe für ihre journalistische Arbeit.*

Ein t-Test bei einer Stichprobe mit dem Trennwert 2 ergibt eine mittlere Differenz von -0,216. $H_0$ wird beibehalten (t(307) = -3,21, p < ,001).

Die Dokumentation des Rechercheprozesses wird von 36,3 Prozent der Journalistinnen und Journalisten nie oder selten ausführlich durchgeführt. Ein Viertel beantwortet die Frage mit „gelegentlich". 38 Prozent dokumentieren oft oder immer ausführlich.

*$H_0$: Die Mehrheit dokumentiert ihren Rechercheprozess gelegentlich, oft oder immer ausführlich.*

*$H_1$: Die Mehrheit dokumentiert ihren Rechercheprozess nie oder selten ausführlich.*

Ein t-Test bei einer Stichprobe mit dem Trennwert 2 ergibt eine mittlere Differenz von 1,108. $H_0$ wird beibehalten (t(307) = 15,97, p < ,001).

Analyse-Tools werden bei einem Mittelwert von 2,6 (SD = 1,366) eher gelegentlich von Journalistinnen und Journalisten zur eigenen Auswertung ihrer journalistischen Veröffentlichungen genutzt. 52,2 Prozent nutzen solche Software nie oder selten selbst. 25,7 Prozent oft oder immer.

*$H_0$: Die Mehrheit nutzt gelegentlich, oft oder immer Analyse-Tools, um Kennzahlen Ihrer journalistischen Veröffentlichungen selbst auszuwerten.*

*$H_1$: Die Mehrheit nutzt nie oder selten Analyse-Tools, um Kennzahlen ihrer journalistischen Veröffentlichungen selbst auszuwerten.*

Ein t-Test bei einer Stichprobe mit dem Trennwert 2 ergibt eine mittlere Differenz von 0,604. $H_0$ wird beibehalten (t(307) = 7,76, p < ,001).

Zwei Drittel der Journalistinnen und Journalisten würde in Zukunft umfangreiche Angaben zum Rechercheprozess machen, wenn eine Plattform diesen Prozess vereinfachen würde. 14,3 Prozent lehnen dies ab und 19,6 Prozent sind unentschlossen. Ein Analyse-Dashboard zur Auswertung der eigenen Veröffentlichung würden 72 Prozent der Journalistinnen und Journalisten nutzen. Software als Produktionshilfe würden – auch bei besserer Qualität – 38,2 Prozent nicht oder eher nicht nutzen.

Tab. 7: Hypothesentests zur Nutzungsintention von Software als Produktionshilfe

| $H_0$ | Die Mehrheit würde Software als Produktionshilfe in Zukunft nicht häufiger nutzen. | |
|---|---|---|
| $H_{1a}$ | Die Mehrheit würde Software als Produktionshilfe in Zukunft nutzen, insofern die Qualität besser würde. | $t(308) = 0{,}268$, $p > {,}05$ $H_0$ beibehalten |
| $H_{1b}$ | Die Mehrheit würde in Zukunft umfangreiche Angaben zum Rechercheprozess machen, wenn eine Plattform dies vereinfachen würde. | $t(308) = 12{,}694$, $p < {,}001$ $H_1$ vorläufig angenommen |
| $H_{1c}$ | Die Mehrheit würde in Zukunft ein speziell zugeschnittenes Analyse-Dashboard nutzen. | $t(308) = 15{,}014$, $p < {,}001$ $H_1$ vorläufig angenommen |

# 5 Methodische und inhaltliche Limitationen

Herstellung von Repräsentativität ist die zentrale methodische Herausforderung für Befragungen von Journalistinnen und Journalisten in Deutschland, aufgrund einer unklaren Berufsbezeichnung sowie daraus resultierender Unsicherheiten über die Grundgesamtheit (Malik 2011). Die durchgeführte Fallgewichtung führt durch das einfache Multiplizieren der einzelnen Gewichtungsvariablen nicht zur mathematisch korrekten Verteilung. Komplexere Fallgewichtungsprozeduren wie die von Gabler und Ganninger (2010) sollten angewendet werden.

Trotz Pretest und ausführlicher Erklärung können die Autoren davon ausgehen, dass das Konzept der Klubfinanzierung nicht von allen Befragten verstanden wurde. 20 Freitextantworten umschreiben das vorgestellte Konzept mit anderen Worten (siehe Kapitel 4.2).

Bei der Abfrage des Einkommens wurde die Kategorie „3500-4000 Euro" vergessen. Da nicht bestimmt werden kann, inwiefern entsprechend verdienende Umfrageteilnehmer, geantwortet haben, wird die Auswertung des Einkommens von der Analyse ausgeschlossen.

# 6 Fazit

Zwei Drittel der Journalistinnen und Journalisten in Deutschland haben die Bereitschaft eine Journalismusplattform – also eine sektorale Plattform der Journalismus-Branche – hauptberuflich zu nutzen, insofern diese den eigenen Lebensunterhalt sichert. Lediglich jede zehnte Journalistin und jeder zehnte Journalist hat weder eine haupt- noch nebenberufliche Nutzungsbereitschaft für eine solche Plattform.

Die Mehrheit der Journalistinnen und Journalisten sieht das monatliche Abonnement einer Plattform als attraktivstes Finanzierungsmodell, gefolgt von einer Klubfinanzierung – also einer Finanzierung durch ein monatliches Abonnement eigener Follower. Ein Werbeerlösmodell wird im Vergleich 20 Prozent weniger attraktiv bewertet. Ähnlich zur Werbung werden im Mittel Zahlungen für einzelne Veröffentlichungen (sog. Micro-Payments) als neutral gesehen, wobei die Antworten eine bimodale Verteilung aufweisen und somit auf starke Meinungsdivergenzen hinweisen.

Den Journalistinnen und Journalisten ist wichtig, dass eine Journalismusplattform in Deutschland oder Europa betrieben wird, dass sie besonderen Wert auf Datenschutz legt und unabhängig von etablierten Tech-Giganten wie Google und Facebook ist. Neben diesen Rahmenbedingungen einer Plattform, sollen Veröffentlichungen der Befragten überwiegend mit ihrem eigenen Namen statt mit einer Redaktionsmarke dargestellt werden. Ein One-Stop-Shop für journalistische Inhalte wird gegenüber mehrerer Insellösungen bevorzugt. Die transparente Darstellung der Auswahlentscheidungen von Empfehlungsalgorithmen sehen Journalistinnen und Journalisten als klaren Vorteil für Rezipierende. Dagegen wird eine algorithmische Qualitätsbewertung einzelner Veröffentlichungen als Eingriff in die journalistische Freiheit betrachtet. Bei den Fragen nach einer personalisierten Startseite und einer generellen Vorauswahl von „Qualitätsmedien" divergieren die Meinungen der Befragten stark.

Die Mehrheit der Journalistinnen und Journalisten nutzt multimediale Darstellungsformen und interaktive Elemente gelegentlich, oft oder immer und intendiert dies in Zukunft häufiger zu tun. Aneinandergereihte Bilder und Videos (sog. Storys oder Reels) werden selten oder nie genutzt, ihre Nutzung in Zukunft aber häufiger intendiert.

Die Mehrheit der Journalistinnen und Journalisten nutzt nie oder selten Software auf Basis „Künstlicher Intelligenz" als Produktionshilfe oder Analyse-Tools, um selbst Kennzahlen ihrer Veröffentlichungen auszuwerten. Darüber hinaus dokumentieren die meisten ihren Rechercheprozess nie, selten oder gelegentlich ausführlich. Ein persönliches Analyse-Dashboard sowie eine technische Möglichkeit umfangreiche Angaben zum Rechercheprozess zu machen, würden Journalistinnen und Journalisten in Zukunft nutzen. Software als Produktionshilfe für Überschriftenvorschläge, Textbausteine, eine automatische Bildauswahl oder die Optimierung von Text würden die meisten auch in Zukunft, insofern die Qualität für die deutsche Sprache besser würde, nicht nutzen.

Bislang existiert kein konkretes Umsetzungsbeispiel einer Journalismusplattform, die vielfältigen und qualitativen Journalismus anbietet, transparente Personalisierungsalgorithmen nutzt, ein Bezahlmodell hat, welches Gesellschaftsinteresse und Nutzerinteresse vereinen kann und in Folge Journalismus aus theoretischer, medienmeritorischer Perspektive für das Internet denkt. Daraus resultieren für die empirische Forschung methodische Limitationen, beispielsweise bei der Ermittlung von Zahlungsbereitschaft für digitalen Journalismus, da entsprechende Befragungen in die Zukunft gerichtete, hypothetische Fragen verwenden müssen. "Die Erwartungen der Konsumentinnen und Konsumenten orientieren sich dabei auffallend stark an den Charakteristika etablierter Plattformen wie Netflix und Spotify" (Buschow und Wellbrock 2020, S. 127). Für die weitere quantitative Analyse der „Plattform-Frage" sollten künftige Forschungsarbeiten mit Rezipientenfokus das wissenschaftliche Experiment wählen. Ebenfalls sollten in Gruppendiskussionen zu den Zukunftsperspektiven des digitalen Plattform-Journalismus bisher undokumentierte Phänomene mit ihren ursächlichen Bedingungen, Kontexten, Konsequenzen und Strategien in der sogenannten Generation Y und der Generation Z erforscht werden.

Der schwedische Soziologe Jan Ekecrantz schreibt bereits 2006: "die Zukunft des Journalismus [..] wird gleichzeitig global, konfliktgeladen, digital, online und mobil, privat und individuell, interaktiv, visuell, performativ und fiktional - und demzufolge post-journalistisch und post-mainstream sein - dann kristallisieren sich mit diesen Begriffen zehn mögliche Folgen für die geopolitischen, technologischen, sozialen, ästhetischen und journalistischen Gegebenheiten heraus." (Ekecrantz 2006)

Die Autoren der Studie unterstützen diese bald zwanzigjährige Zukunftsbeschreibung, konkretisieren sie an dieser Stelle um zwei Thesen: Unabhängiger digitaler Journalismus braucht ein Meritorik-Konzept und Mehrwertkommunikation, um die „Gratis-Mentalität" in Deutschland zu überwinden. Nicht die Zahlungsbereitschaft ist das Problem des digitalen Journalismus, sondern das Produkt für das gezahlt werden soll.

# 7 Literaturverzeichnis

# 8 Anhang

**Berechnung der Stichprobengröße**

$$\text{Stichprobengröße} = \frac{\frac{z^2 \times p(1-p)}{e^2}}{1 + \left(\frac{z^2 \times p(1-p)}{e^2 N}\right)}$$

**Berechnung der Fallgewichtung**

COMPUTE genwt = 61/53*(sex = 1) + 39/47*(sex = 2).

COMPUTE freewt = 61/65*(anstellung = 0) + 39/35*(anstellung = 1).

COMPUTE agewt = 8.4/27.2*(agegrp = 1) + 19.8/26.6*(agegrp = 2) + 16.8/21.3*(agegrp = 3) + 55/25*(agegrp = 4).

COMPUTE wt = genwt* freewt* agewt.

WEIGHT BY wt.